\documentclass[11pt,twoside]{article}
\usepackage{./asp2014}

\aspSuppressVolSlug
\resetcounters

\def\msol{\hbox{$\hbox{M}_\odot$}}
\newcommand   {\kms}   {\mbox{km\,s$^{-1}$}}
\def\msol{\hbox{$\hbox{M}_\odot$}}

\begin{document}

\title{Cosmic-ray Particles in the Galactic Center: Blowing in the Wind}
\author{F. Yusef-Zadeh$^1$, M. Wardle$^2$, I. Heywood$^{3,4,5}$, W. Cotton$^6$, and M. Royster$^1$} 
\affil{$^1$Northwestern University, Evanston, Illinois, USA; \email{zadeh@northwestern.edu}}
\affil{$^2$Macquarie University, Sydney, NSW, Australia; \email{mark.wardle@mq.edu.au}}
\affil{$^{3}$Astrophysics, Department of Physics, University of Oxford, Keble Road, Oxford, OX1 3RH, UK; \email{ian.heywood@physics.ox.ac.uk}}
\affil{$^{4}$Department of Physics and Electronics, Rhodes University, PO Box 94, Makhanda, 6140, South Africa}
\affil{$^{5}$South African Radio Astronomical Observatory, 2 Fir Street, Black River Park, Observatory, Cape Town, 7925, South Africa}
\affil{$^6$National Radio Astronomy Observatory, Charlottesville, VA, USA; \email{bcotton@nrao.edu}}

% This section is for ADS Processing.  There must be one line per author.
\paperauthor{Sample~Author1}{Author1Email@email.edu}{ORCID_Or_Blank}{Author1 Institution}{Author1 Department}{City}{State/Province}{Postal Code}{Country}
\paperauthor{Sample~Author2}{Author2Email@email.edu}{ORCID_Or_Blank}{Author2 Institution}{Author2 Department}{City}{State/Province}{Postal Code}{Country}
\paperauthor{Sample~Author3}{Author3Email@email.edu}{ORCID_Or_Blank}{Author3 Institution}{Author3 Department}{City}{State/Province}{Postal Code}{Country}

\begin{abstract}
Recent results from multi-wavelength observations of the inner few hundred pc of the 
Galactic center have added two new characteristics to the ISM in this unique region.
One is the cosmic ray ionization rate derived from H$_3^+$ measurements is at least two orders
of magnitudes higher than in the disk of the Galaxy.  The other is the 
bipolar thermal X-ray and synchrotron emission from this region, suggesting a relic of
past activity. 
We propose that 
the  high cosmic ray pressure  drives a  large-scale wind away from the
Galactic plane and produces the bipolar emission as well as 
highly blue-shifted diffuse gas detected in  H$_3^+$ absorption studies.
We then discuss the interaction of large-scale winds with a number of objects,  such as
cloudlets and stellar wind bubbles,  to  explain the unusual characteristics of the ISM in this region including 
the nonthermal radio filaments.
One of the implications of this scenario is the removal of gas
driven by outflowing winds may  regulate  star formation or black
hole accretion. 
\end{abstract}
{\noindent\bf\center{Evidence for High Cosmic Ray Flux in the Galactic Center}} 

%This proposed picture suggests that the nucleus of our Milky Way has experienced the
%same type of activities found in the nuclei of disk galaxies either due to starburst
%or black hole activity. 

%\section{Evidence for High Cosmic Ray Flux in the Galactic Center}

It is now becoming clear from multi-wavelength observations that cosmic ray flux is high in the inner few hundred pc of the 
Galaxy. A full array 
of effects come into play in the rich environment of the Galactic center when we consider the interacting cosmic rays with 
the ISM, the magnetic fields and stellar winds.  The most direct and accurate estimate of high cosmic ray flux comes from 
H$_3^+$ studies because of its simple production pathway through cosmic ray ionization of 
molecular hydrogen.  
Detailed modeling yields ionization rates 
$\zeta\sim2\times10^{-14}$ s$^{-1}$ (Oka et al. 2019) or  $(1-11)\times10^{-14}$ s$^{-1}$ (LePetit et al. 2016). H$_3^+$ 
spectra also show that the  gas has to be warm (T$\sim$200 K), diffuse with density 50--100 cm$^{-3}$  and   a total mass 
of  $\sim6\times10^6$ \msol\,  in its atomic phase.  
H$_3^+$ absorption profiles are highly blue-shifted, and show broad linewidth that 
extends between --150 and 0 \kms\,  with a mean momentum of $\sim2-6\times10^8$ \msol\, \kms. 

%The idea  of a diffuse warm component filling a significant fraction of the CMZ runs counter to the traditional view of a 
%CMZ dominated by dense molecular gas. 

%A significant population of H$_3^+$ in the (3,3) level is now recognized to be the $``$fingerprint'' of 
%previously unknown  diffuse gas pervading  the Central Molecular Zone, CMZ (Oka et al. 2019). 

%H$_3^+$  measurements imply that dense gas in the CMZ 
%has a low volume filling factor much less than 0.1 (Oka et al. 2019). 

There is  strong  nonthermal radio continuum emission from the CMZ, including  numerous magnetized filamentary structures.
Thus, nonthermal radio continuum emission can also be  used to probe the population of cosmic-ray electrons, 
and independently used to 
a high (albeit more uncertain) estimate 
of  $\zeta$ in the CMZ (Yusef-Zadeh et al. 2013). 
Now that observational studies 
of H$_3^+$ absorption toward 30 stellar sources in the CMZ cemented the idea that the Galactic center H$_2$ 
cosmic ray ionization rate  is higher than in diffuse or dense clouds in the Galactic disk by two or three orders of magnitudes, 
respectively 
we explore some of the 
cosmic ray effects in this region. 

%(Geballe et al. 1999; Indriolo et al. 2012; Oka et al. 2005, 2019), 

{\noindent\bf\center The 6.4 keV line and  $\gamma$-ray continuum emission:} 
There is pervasive, steady  detection of the 6.4 keV line emission
from the neutral iron line emission tracing  cold neutral   gas in the CMZ. Remarkably Suzaku measurements have shown 
that the spatial distribution of K$\alpha$ line emission is remarkably similar to molecular line emission.  
The origin of the Fe 6.4 keV line emission 
is  explained by irradiation of molecular clouds by X-rays from an energetic burst from Sgr A* a few hundred years ago 
 or the impact of low energy cosmic ray electrons  and protons (Ponti et al. 
2010; Yusef-Zadeh et al. 2007). 
Nonthermal radio continuum emission is used to probe the population of cosmic-ray electrons. 
The low energy  component of these cosmic-ray electrons and protons  interact with 
the reservoir of molecular gas distributed in the Galactic center and produce 
the 6.4 keV line emission. 
The intensity of K$\alpha$  
photons from molecular clouds  subject to a cosmic-ray-particles-induced ionization rate per hydrogen nucleus is estimated to be 
similar to that found from H$_3^+$ measurements. 
A vast volume of literature exits   arguing  
against the role of cosmic rays in the origin of 
the 6.4 keV emission since the role of cosmic rays  was suggested nearly  two decades ago (Yusef-Zadeh et al. 2002).  
It is now becoming apparent 
that bombardment of  molecular clouds by low energy cosmic rays is a viable mechanism in producing the steady component of the 
6.4 keV line emission. 

We have also shown  that the emission detected by Fermi is primarily due to 
nonthermal bremsstrahlung produced 
by the population of synchrotron emitting electrons in the GeV energy range interacting with neutral gas.
Very high energy $\gamma$-ray emission detected by HESS indicated a correlation between 
high energy cosmic rays and the  molecular gas in the CMZ. 
$\gamma$-ray  measurements  toward the CMZ infer a sea of high energy cosmic ray particles with 
a high value of cosmic ray energy density compared to the Galactic disk 
(HESS collaboration 2017). 

{\noindent\bf\center Launching cosmic-ray driven winds:} 
Two recent radio  and X-ray observations indicate 
high cosmic ray flux away from the Galactic plane where the CMZ lies. 
Assuming that synchrotron emission is a  proxy for relativistic cosmic ray particles, 
 MeerKAT observations of the Galactic center 
discovered a bubble of synchrotron emission  spanning over 430 pc above and below the plane (Heywood et al. 2019). 
The  bubble appears to be filled with hot coronal X-ray gas 
indicating that an energetic outflow took place few times $10^{5-6}$ years ago (Nakashima et  al. 2019; Ponti et al. 2019; Heywood et al. 2019). 
The cosmic ray driven outflow  will expand at its effective sound speed  750 \kms\, with 
an estimate outflow rate   $\dot{M} \approx 0.075$\,\msol\,yr$^{-1}$,
with a radius of  $r\approx75$\,pc (Yusef-Zadeh and Wardle 2019). 

%This is consistent with past studies indicating that  cosmic rays driven  
%winds in the nuclei of galaxies (e.g., Everett et al. 2008, 2010; Ruskowski et al. 2017; Zweibel 2017) but this is the first time
% that this idea been applied to the CMZ. 

{\noindent\bf\center Warm gas in the CMZ:} 
Numerous studies have indicated that the gas temperature of molecular clouds in the CMZ is
high. This is consistent with H$_3^+$ absorption studies indicating warm gas in the CMZ. A global heating mechanism is needed 
to explain the presence of a pervasive warm medium in the CMZ. We showed that the cosmic ray heating is consistent with the 
ionization rate, thus naturally explains why the gas temperature is $\sim200$K and why it is significantly higher than the 
dust temperature (Yusef-Zadeh et al. 2007). It is also possible that turbulent heating is responsible for elevating  the gas 
temperature but the decay time scale of turbulence is shorter than the cooling times scale, thus has to be replenished.

{\noindent\bf\center Highly blue-shifted diffuse atomic gas:} 
It is possible that the highly blue-shifted diffuse medium was  originated by the erosion of dense clouds 
by cosmic ray dissociation and heating and then accelerated to high velocities by the  Galactic center wind. 
In this picture, the evaporation of the surface 
layer of clouds produce an atomic medium that can then be pushed away in the form of a cosmic-ray driven wind, producing the 
highly blue-shifted diffuse gas traced from H$_3^+$ measurements.  
The broad blue-shifted diffuse atomic gas is estimated to have  a momentum of $3\times10^8$ \msol\, \kms 
(Oka et al. 2019 and Geballe in this conference).  The cosmic ray outflow pushing the gas isotropically  must 
accelerate the diffuse  gas for  $\sim5\times10^6$ years. 

%and be responsible for the observed blue-shifted atomic gas.  

{\noindent\bf\center Vertical  and azimuthal magnetic fields:} 
If the magnetic field is in equipartition with cosmic ray particles, the magnetic field strength would be very large in the CMZ. 
However, the assumption is that the injection of cosmic rays took place by a transient  event implying that  
the equipartition argument between cosmic ray particles  and the magnetic field 
can not be applied.   The magnetic field 
strength is estimated to be $\sim4.3 \mu$G assuming the 
observed synchrotron intensity and cosmic ray electron density (Yusef-Zadeh and Wardle 2019). The wind can easily push the azimuthal magnetic field and distort the field lines to be vertical away from the plane. 

{\noindent\bf\center Magnetized radio filaments:} 
Radio observations have identified a system of magnetized filamentary structures 
in  the inner two degrees of the Galactic center and vertical lobes at the edges of the bubble (Heywood et al. 2019). 
The large-scale wind away from the Galactic plane 
will be interacting with a variety of  
objects that act  as obstacles. These  include  compact and extended HII regions, stellar bubbles, pulsar wind nebulae or planetary nebulae. 
In this picture, the weak magnetic field in the wind wraps around the
 obstacle  and forms a long tail behind the star.
The outflowing Galactic center wind and its magnetic field are directed
vertically away from the plane, but the horizontal component of the
 orbital velocity of the object means that magnetic field lines become draped
over the obstacles. The tail must  naturally have a higher magnetic field because 
the magnetized wind are  pressed together behind the obstacle. 

A large number of  nonthermal filaments and diffuse ionized thermal gas  
are detected in the radio lobes at the edge of the MeerKAT bubble. 
Since the hot X-ray gas and large-scale outflow  are  overpressured with respect to the exterior, 
the expanding gas sweeps low density ambient gas and 
the magnetic fields to the sides and forms  a region where  
magnetic field lines become stronger  with higher  volume filling factor. 
% of the ionized gas becomes higher. 

{\noindent\bf\center High velocity HI clouds:} 
Large-scale Galactic winds can also carry dense clouds to high altitudes. A number of high velocity 
compact HI clouds are found away from the Galactic plane and interpreted as entrained gas clouds in the outflowing winds
associated with Fermi bubbles (McClure-Griffiths et al. 2013).
We note that some of the low-latitude HI clouds lie close to the southern bubble discovered by 
MeerKAT.  Figure 1a,b show an HI channel map  of a coherent balloon-like structure at a velocity of -90.7  \kms 
and a 20cm continuum image of the  MeerKAT bubble, respectively.  
The velocity structure of this elongated HI  balloon
shows highest negative velocity of $\sim-100$  \kms  to the north and  becoming less negative to the south 
with a velocity of -73 \kms.  
It is not clear if the MeerKAT bubble and the HI balloon are associated with each other until additional observations are made. 
The strongest HI emission  coincides a compact cloud at the southern edge of the HI balloon 
at the position
l=359$^0$.39 and b=-3$^0$.38 with the brightness temperature 7.6$^0$K 
(source number 80 in Table 1 of  McClure-Griffiths et al. 2013). 
This HI cloud has  a  column density of 2.6$\times10^{20}$ cm$^{-2}$ and a total HI gas of 477 \msol. 

%chemistry

\articlefiguretwo{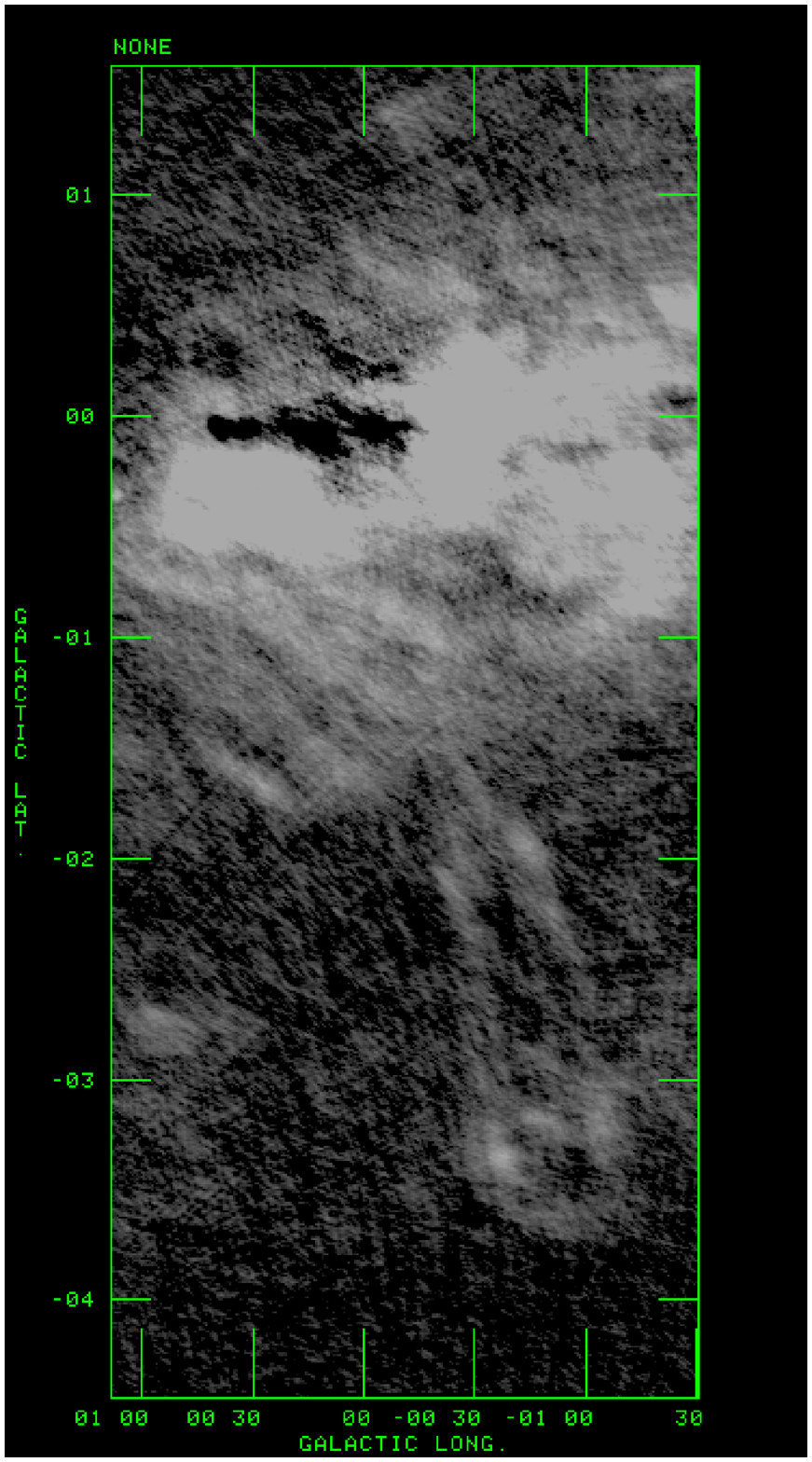}{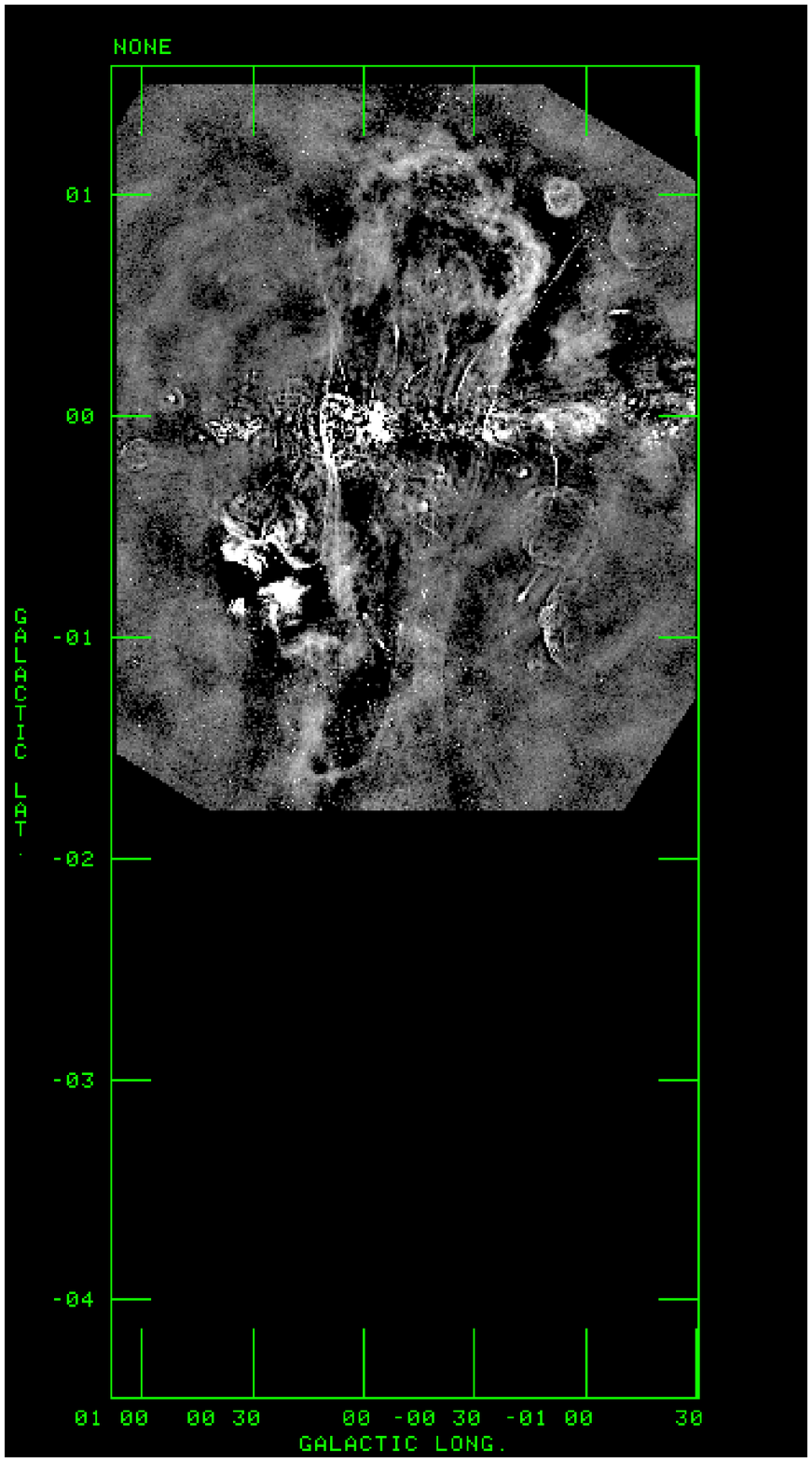}{ex_fig2}{(a Left) HI  image at -90.7 \kms\, from ATCA (McClure-Griffiths et al. 2013).  (b Right) The same region as (a), 20cm continuum  image from MeerKAT (Heywood et al. 2019).}

%\articlefiguretwo[width=.5\textwidth]{atca_ch266HI.eps}{meerkat_ch266HI.eps}{ex_fig2}{Now there are two of them. The same exact thi}

\end{document}